\documentclass[aps,prl,
preprint,
superscriptaddress,
]{revtex4}

\usepackage[pdftex]{graphicx}
\usepackage{amssymb}
\usepackage{amsmath,amsfonts,latexsym}
\usepackage{array,tabularx,color}
\usepackage{dcolumn}               
\usepackage[normalem]{ulem}
\usepackage{comment}
\usepackage[amssymb]{SIunits}

\newcommand{\vect}[1]{{\mathbf #1}}

\begin{document}

\title{Persistent currents and quantised vortices in a polariton superfluid}

\author{D. Sanvitto}
\email{daniele.sanvitto@uam.es}
\affiliation{Departamento de F\'isica de Materiales, Universidad
  Aut\'onoma de Madrid, Madrid 28049, Spain}

\author{F. M. Marchetti}
\email{francesca.marchetti@uam.es}
\affiliation{Departamento de F\'isica Te\'orica de la Materia
  Condensada, Universidad Aut\'onoma de Madrid, Madrid 28049, Spain}

\author{M. H. Szyma\'nska}
\affiliation{Department of Physics, University of Warwick, Coventry,
  CV4 7AL, UK}

\author{G. Tosi}
\affiliation{Departamento de F\'isica de Materiales, Universidad
  Aut\'onoma de Madrid, Madrid 28049, Spain}

\author{M. Baudisch}
\affiliation{Departamento de F\'isica de Materiales, Universidad
  Aut\'onoma de Madrid, Madrid 28049, Spain}

\author{F. P. Laussy}
\affiliation{School of Physics and Astronomy, University of Southampton, Southampton, SO17 1BJ, UK}

\author{D. N. Krizhanovskii}
\affiliation{Department of Physics and Astronomy, University of
Sheffield, Sheffield, S3 7RH, UK}

\author{M. S. Skolnick}
\affiliation{Department of Physics and Astronomy, University of
Sheffield, Sheffield, S3 7RH, UK}

\author{L. Marrucci}
\affiliation{Dipartimento di Scienze Fisiche, Universit\`a di
Napoli Federico II and CNR-SPIN, Napoli, Italy}

\author{A. Lema\^{i}tre}
\affiliation{LPN/CNRS, Route de Nozay, 91460, Marcoussis, France}

\author{J. Bloch}
\affiliation{LPN/CNRS, Route de Nozay, 91460, Marcoussis, France}

\author{C. Tejedor}
\affiliation{Departamento de F\'isica Te\'orica de la Materia
  Condensada, Universidad Aut\'onoma de Madrid, Madrid 28049, Spain}

\author{L. Vi\~na}
\affiliation{Departamento de F\'isica de Materiales, Universidad
  Aut\'onoma de Madrid, Madrid 28049, Spain}


\date{\today}













\maketitle

\textbf{After the discovery of zero viscosity in liquid helium, other
fundamental properties of the superfluidity phenomenon have been
revealed. One of them, irrotational flow, gives rise to quantised
vortices and persistent currents. Those are the landmarks of
superfluidity in its modern understanding. Recently, a new variety of
dissipationless fluid behaviour has been found in microcavities under
optical parametric regime. Here we report the observation of
metastable persistent polariton superflows sustaining a quantised
angular momentum, $m$, after applying a
2 ps pulsed probe carrying a
vortex state. We observe a transfer of angular momentum to the
condensate steady state, which sustains vorticity for as long as it
can be tracked. Furthermore, we study the stability of quantised
vortices with $m=2$. The experiments are analysed via a
generalised two-component Gross-Pitaevskii equation. These results,
aside from demonstrating the control of metastable persistent currents,
show the peculiar superfluid character of non-equilibrium polariton
condensates.}

In the past few decades there has been a strenuous search for
macroscopic coherence and phenomena related to Bose-Einstein
condensation (BEC) in the solid state. The first realisation of a BEC
in semiconductor microcavities~\cite{kasprzak06:nature} has
inaugurated a new era in the study of strongly coupled light-matter
systems. The growing interest in this field can be attributed to the
unique properties of exciton-polaritons in
microcavities~\cite{weisbuch92}, the composite particles resulting
from strong light-matter coupling.  The properties of a polariton
condensate~\cite{keeling_review07} differ from those of other known
condensates, such as ultracold atomic BECs and superfluid $^4$He. In
particular, polaritons have a short lifetime of the order of
picoseconds, therefore needing continuous pumping to balance decay and
reach a steady state regime. Rather than a drawback, the intrinsic
non-equilibrium nature enriches the features of polariton
condensation, but at the same time poses fundamental questions about
the robustness of the coherence phenomena to dissipation and
non-equilibrium. Superfluid properties of non-equilibrium condensates
in dissipative environment still need to be
understood~\cite{keeling2009}. In this work, to advance in this
direction, we investigate the hallmark of superfluidity, namely,
vortices and metastable persistent flows.

\section*{Polariton superfluid phases out of equilibrium}
%
One route to inject polaritons into a microcavity is by non-resonant
(incoherent) pumping. For incoherent pumping, polaritons have been
shown to enter, within their short lifetime, a macroscopically
coherent BEC
phase~\cite{kasprzak06:nature,snoke07science,lai07a}. However, the
unusual form of the excitation spectrum---diffusive at small
momenta---hinders the fulfilling of the Landau criterion and puts
under debate the possibility of dissipationless superflow in
incoherently pumped polariton
systems~\cite{szymanska06:prl,wouters:140402,wouters10}. Because of the
non-equilibrium nature of the polariton condensate in an inhomogeneous
system, there are spontaneous supercurrents that may carry polaritons
from gain to loss dominated regions. This can give rise to spontaneous
formation of deterministic vortices, that do not necessarily imply
superfluidity~\cite{lagoudakis08,keeling2009}. Another peculiarity of
polariton systems is to be found in their polarization, giving rise to
a nomenclature of half-vortices~\cite{Rubo07,lagoudakis09a}.

A different scenario characterises coherent resonant injection of
parametrically pumped polariton condensates, which have been recently
shown to exhibit a new form of non-equilibrium superfluid
behaviour~\cite{amo2009,amo2009_b}. In the optical parametric
oscillator (OPO) regime~\cite{stevenson00}, bosonic final state
stimulation causes polariton pairs to coherently scatter from the pump
state to the signal and idler states, which, at threshold, have a
state occupancy of order one. The properties of the quantum fluids
generated by OPO at idler and signal have been recently tested via a
triggered optical parametric oscillator (TOPO)
configuration~\cite{amo2009}. An additional weak pulsed probe laser
beam has been used to create a traveling, long-living, coherent
polaritons signal, continuously fed by the OPO. The traveling signal
has been shown to display superfluid behaviour through frictionless
flow. However, long-lived quantised vortices and metastable persistent
flow, i.e., another possible steady state of the condensate, still
remained missing in the superfluid `checklist'~\cite{keeling2009}. In
this work, we fill-in this gap and we also address the case of higher
winding numbers, unravelling a rich dynamics of vortices in polariton
condensates.
For this purpose, we use a technique already applied in non-linear and
quantum optics, cold atoms and biophysics~\cite{molina}: excitation by
a light field carrying orbital angular momentum (OAM). Transfer of
light OAM has been demonstrated in parametric processes in non-linear
materials~\cite{dholakia,martinelli} and has been used to generate
atomic vortex states in Bose-Einstein condensates~\cite{andersen06,ryu}.

\section*{Generation of persistent currents}
\begin{figure}[htbp]
\begin{center}
\includegraphics[width=.55\linewidth,angle=0]{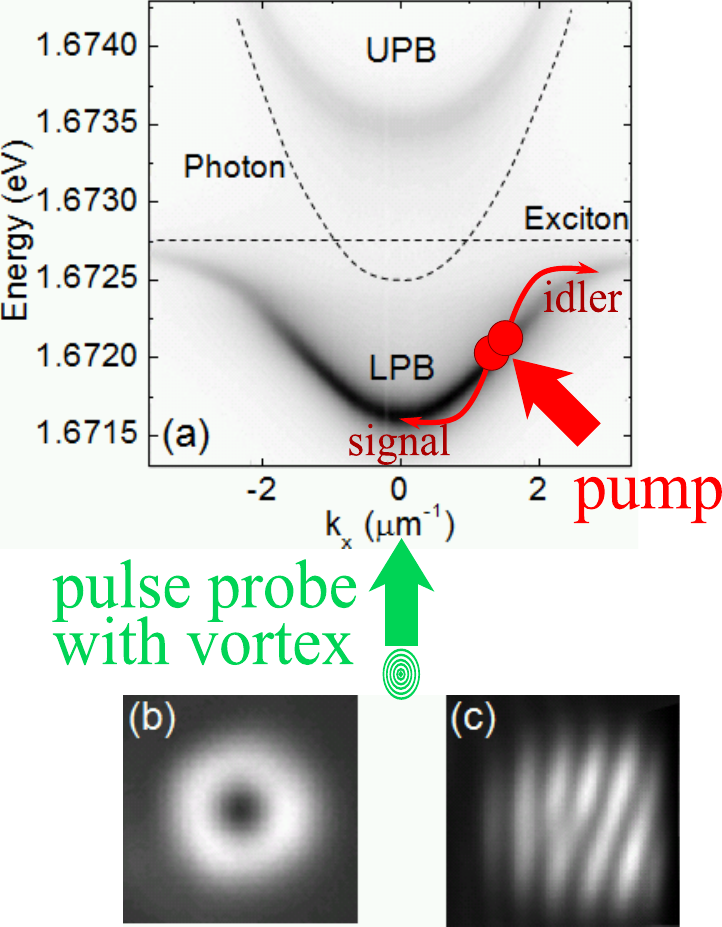}
\end{center}
\caption{Lower (LPB) and upper polariton branch (UPB) dispersions
  together with the schematic representation of the TOPO excitation
  (a). Resonantly pumping the LPB initiates, above threshold,
  stimulated scattering to a signal close to zero momentum and an
  idler at higher momentum. The $m=1$ Laguerre-Gauss beam resonant
  with the signal (b) is used as a weak pulsed triggering probe to
  stir the superfluid. The corresponding interference image is shown
  in panel (c).}
\label{fig:schem}
\end{figure}

\begin{figure}[htbp]
\begin{center}
\includegraphics[width=.93\linewidth,angle=0]{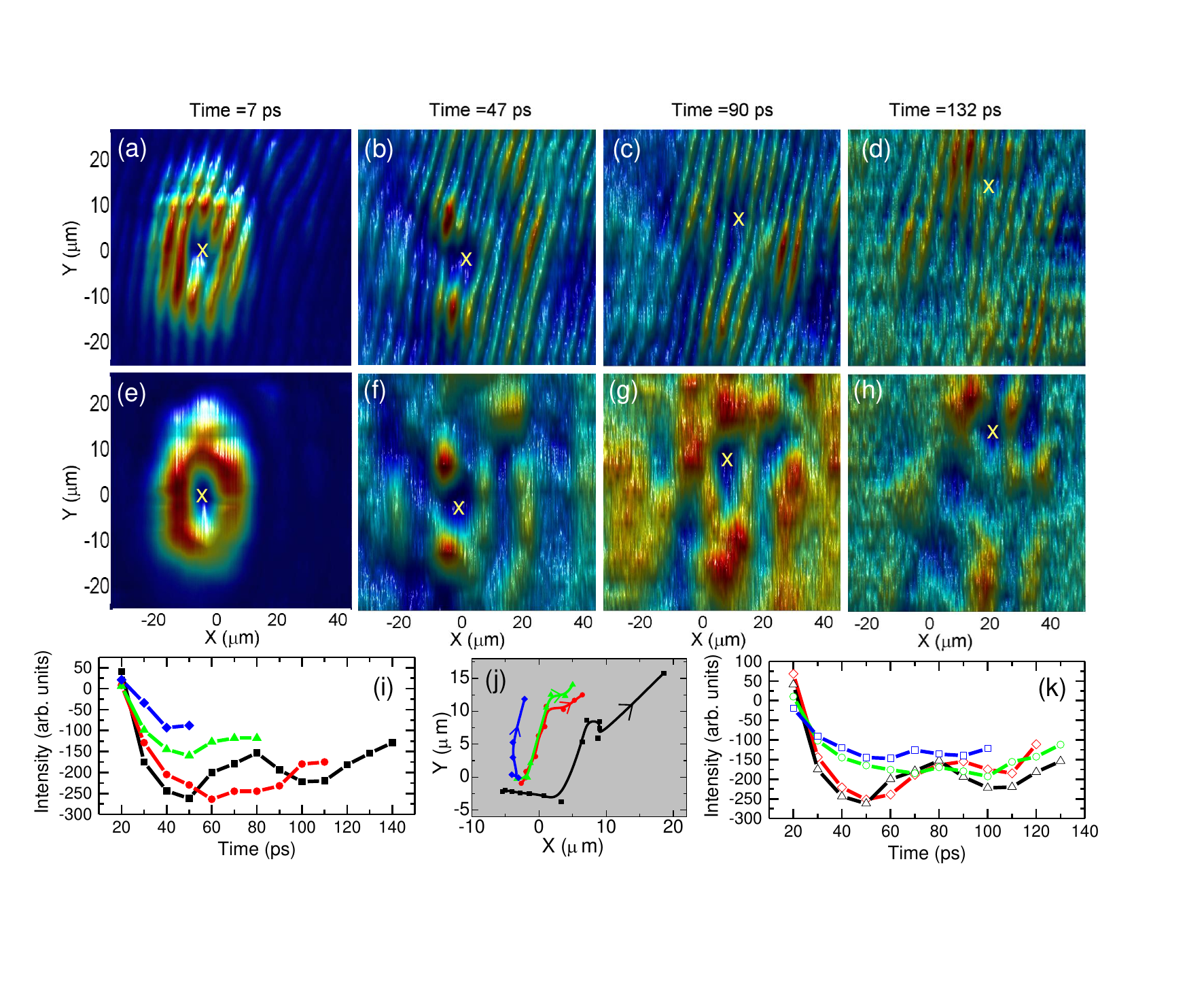}
\end{center}
\caption{Time evolution of the polariton signal after a weak pulsed
  probe with a vortex of $m=1$ has been excited (a-h). The
  interference images (a-d) are obtained by overlapping the vortex
  with a small expanded region of the same image far from the vortex
  core, where the phase is constant. The time origin is taken when the
  extra population has reached $80\%$ of its maximum value. To better
  reveal the effect of the imprinting of the vortex into the
  condensate steady state of the signal, the contribution of the
  unperturbed polariton signal (in absence of the probe pulse) is
  subtracted from all data. The depth of the vortex core and its
  position as a function of time for 4 different pump powers
  (\unit{65}\milli\watt, blue diamonds, \unit{100}\milli\watt, green
  triangles, \unit{200}\milli\watt\ red circles,
  \unit{300}\milli\watt, black squares)---all above the OPO threshold
  (\unit{50}\milli\watt)---is shown in (i) and (j) respectively. The
  effect of probe power---\unit{0.15}\micro\watt\ in blue squares,
  \unit{0.33}\micro\watt\ in green circles, \unit{6.65}\micro\watt\ in
  black triangles, \unit{8}\micro\watt\ in red diamonds---on the
  duration and depth of the core of the vortex is shown in (k).}
\label{fig:persi}
\end{figure}

\begin{figure}[htbp]
\begin{center}
\includegraphics[width=\linewidth,angle=0]{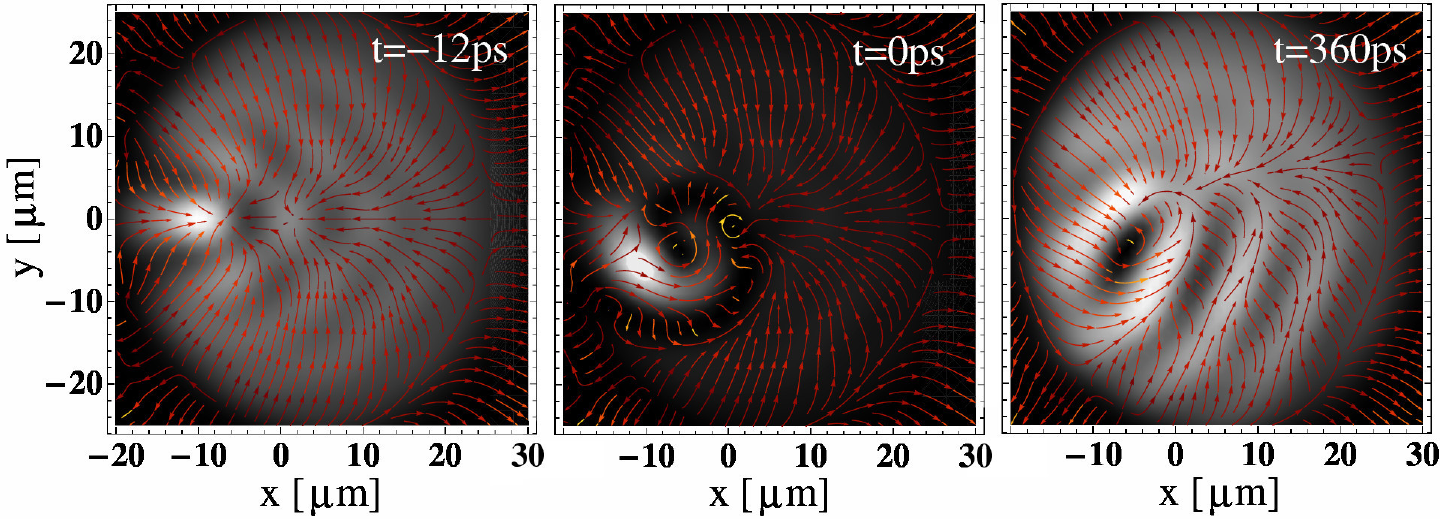}
\end{center}
\caption{Numerical simulations of the time evolution of the OPO signal
  before and after the arrival ($t=0$) of a pulsed probe carrying an
  $m=1$ vortex resonant with the signal momentum and energy, for
  pumping strength $f_p=1.24 f_p^{\text{(th)}}$ above OPO
  threshold. The images of the signal are obtained by momentum
  filtering in a cone of around $\pm 7^{\circ}$. The pulse carrying
  the vortex generates a gain which fades out after about
  \unit{10}\pico\second\ leaving an $m=1$ vortex imprinted into the
  signal. The vortex experiences a transient time of around
  \unit{30}\pico\second\ before settling into a metastable
  solution. The supercurrents are plotted in the frame of the signal
  by subtracting the $\vect{k}$ of the signal.}
\label{fig:thfi1}
\end{figure}

%
In our experiment, the vortex is excited by a pulsed probe resonant
with the signal (see Fig.~\ref{fig:schem}) lasting only
\unit{2}\pico\second. In other words, we stir the polariton superfluid
only for a short time and observe its long lived rotation with a
quantum of angular momentum on a time scale almost 70 times longer
than the duration of the pulse (see Figs.~\ref{fig:persi}
and~\ref{fig:thfi1}). Although there is a similarity with rotating
trapped gases, the effect for driven, non-equilibrium systems shows a
richer phenomenology as will be described in the rest of this paper.
Beside, a characteristic of vortices with higher winding number $m$ is
the tendency to coherently split into many vortices of
$m=1$~\cite{shin2004}. To study the stability properties of vortices
in polariton condensates, we have injected a vortex with $m=2$ and
observed its evolution (see Fig.~\ref{fig:splitt}). Surprisingly, we
found three different behaviours depending on the initial
condition. In cases for which the vortex is imprinted into the signal
steady state, one topological charge is always expelled out of the
condensate (Fig.~\ref{fig:splitt} (g-i)). On the other hand, when the
vortex is not imprinted in the steady state but lasts only as long as
the perturbation of the signal, we observe either a stable vortex of
$m=2$ when the probe carrying the vortex is at small momentum,
resulting in a static (or slowly moving) vortex and slow
supercurrents, or a splitting in two vortices of $m=1$ when the vortex
is injected with a probe at higher momentum.  In the latter case, the
vortex is moving faster than when it does not split, but still slower
than the supercurrents of the condensate. Instability of doubly
quantised vortices in a BEC has been observed in several
systems~\cite{shin2004}, but there are only a few examples of their
stability, such as in superconductors in the presence of pinning
forces~\cite{PhysRevLett.74.3269} or in a multicomponent order
parameter superfluid such as $^3$He-A~\cite{Nature2000}.
In ultra-cold atomic gases, stable free $m=2$ vortices have been
theoretically predicted for some range of densities and interaction
strengths \cite{PhysRevA.68.023611}. However, they have been observed
only in a toroidal pinning potential with an external optical plug
\cite{ryu}, and demonstrated to split soon after the plug has been
removed. The stability of $m=2$ static vortex in polariton systems
provides an additional experimental realization.

We use a semiconductor microcavity with a Rabi splitting
$\Omega_R=\unit{4.4}\milli\electronvolt$ and the cavity photon energy
slightly negatively detuned (between 1 and
\unit{3}\milli\electronvolt) from the exciton energy; see the Method
section for details.  A Ti-Sapphire laser is tuned in resonance with
the lower polariton branch (LPB), injecting polaritons close to the
point of inflection and giving rise to a continuously pumped OPO (see
Fig.~\ref{fig:schem}). Above a pump threshold, the signal generated
close to zero momentum ($\vect{k}_s=0$) as well as the idler state at
high momentum, form an out-of-equilibrium coherent polariton
superfluid. At a given time, we trigger a new scattering process on
top of the OPO signal with a resonant pulsed probe. The size of the
probe is smaller than that of the signal (by a factor $\approx4$) to
allow free motion of the vortex within the condensate, thus avoiding a
spurious confinement. The probe is a pulsed Laguerre-Gauss beam
carrying a vortex of given angular momentum $m$, the phase of which
winds around the vortex core with values from $0$ to $2\pi m$.  After
a short time, $\sim\unit{2}\pico\second$, the probe vanishes, leaving
the polariton coherent state free to rotate, without the driving
field. While even a classical fluid acquires angular momentum in the
presence of an external rotating drive, only a superfluid can exhibit
infinitely lived circulating flow in a dissipative environment once
the external drive is turned off.  In order to demonstrate persistence
of the vortex angular momentum, we detect the phase pattern generated
by making interfere the polariton signal with an expanded and flipped
spatial region far from the vortex core (where the phase is
approximately constant) in a Michelson interferometer. A fork-like
dislocation with a difference of $m$ arms corresponds to phase winding
by $2\pi m$ around the vortex core (see lower panel of
Fig.~\ref{fig:schem}).

%
\section*{Vortex dynamics}
%
Using a streak camera we can follow the evolution in time of the
vortex generated by the pulsed probe. Time intervals of
\unit{4}\pico\second\ are used to reconstruct 2D images of the signal
state after the perturbation has arrived. Every picture is the result
of an average over many shots, all taken at the same time and same
conditions.  In order to separate the contribution of the signal from
that of the pump, we filter, both in theory and experiments, the
signal images in momentum space in a cone around $\vect{k}=0$ of
approximately $\pm 7^{\circ}$. At time $t_{pb}$, the pulsed laser
probe is shined on the sample. Its general effect is to enhance the
polariton signal emission by a factor that ranges, depending on the
experiment, between a few percent to more than 100 \% with a delay of
\unit{10}\pico\second\ after the pulse arrival time. At first a strong
gain given by the presence of the probe creates an extra population on
top of the OPO signal (called TOPO polaritons) before the OPO signal
is re-established. We follow the evolution of the transient state and
monitor how the probe affects the condensate in its steady state. We
find that the original steady state can be recovered but also, more
interestingly, that different stable solutions are possible.

The time evolution of an excited $m=1$ vortex together with its
interference pattern, which characterises unequivocally the vortex
state, are shown in Fig.~\ref{fig:persi}, where the external probe
has a power of $\unit{6.6}\micro\watt$, which is less than the OPO
signal emission before the probe arrives on the sample.  In these
images, we can observe two effects: immediately after the arrival of
the probe, a vortex is generated in the TOPO polaritons and its
vorticity is maintained although the population decays in a few tenths
of picoseconds. After the extra polaritons have disappeared, the
vortex remains imprinted into the steady population of the
condensate. While the polariton vorticity is always present and
enduring in the TOPO polaritons, only under very high pump power and
at specific points in the sample is the vorticity also passed to the
steady state of the OPO signal. This not only demonstrates that
polariton condensates show unperturbed rotation, but also that a
vortex is another stable solution of the final steady state. This is a
clear demonstration of superfluid behaviour in the non-equilibrium
polariton OPO system. These effects are visible in the real space
images through the `toroidal shape' of the polariton emission at early
times, and the presence of a fork dislocation in the interference
images at all times. The latter show that the spatial phase relation
remains constant and with full contrast all over a wide area, even for
times much longer than the signal coherence time, as measured by
detecting the decay of interference-fringe contrast when delaying the
signal and the reference (see Fig.~\ref{fig:visib} in the
supplementary material). This means that even when polaritons have
lost any phase relation in time, their angular momentum is still
conserved.

After the vortex is imprinted into the OPO signal, we can observe the
vortex core slowly drifting to the right and then upwards changing in
shape and moving with different velocities. The drift is due to the
fact that the vortex core is naturally inclined to undergo a random
walk and eventually either remains permanently trapped in the
condensate or is expelled from the edges (see the following theory and
Ref.~\cite{wouters09}). However, due to the presence of local
defects~\cite{sanvitto06}, which tend to influence the motion of the
vortex, the path followed by the vortex core recurs at each repetition
of the experiment. For this reason we are able to track its movement
that otherwise would be washed out by the experimental averaging.

Figures~\ref{fig:persi} (i,j) and (k) show the effect of pump and
probe powers on the transfer of angular momentum to the condensate
steady state. Note that to highlight this effect, all the data shown
in this paper have been obtained by subtracting the steady state of
the condensate unperturbed by the probe pulse. As a consequence, the
depth of the vortex core comes out with negative values. An increase
of the power of the pump helps the observation of this effect possibly
due to an enhancement of the OPO coherence for higher pumping
intensities. Above four times the OPO threshold, we observe a
saturation in the depth of the core [see intensity at
  \unit{50}\pico\second\ in Fig.~\ref{fig:persi} (i)]. However given
the different pathways followed by the vortex when changing pump
intensity [see Fig.~\ref{fig:persi} (j)], a variation of the depth in
time is also expected, depending on the kind of inhomogeneities met by
the vortex along its trajectory. As for the probe, we find that a
minimum power is required for the polaritons to acquire enough angular
momentum to be able to transfer it to the steady state.  This rigidity
of the fluid to accommodate a vortex at low intensities has also been
predicted for incoherently pumped polariton
condensates~\cite{wouters09}. However, once the transfer is achieved,
the probe power does not change significantly the duration and depth
of the vortex in the steady state.
An alternative interpretation of the short vortex lifetime, for low
excitation powers, could be attributed to dissipation when the average
angular velocity is higher than some critical velocity--leading to the
onset of macroscopic drag forces (like recently discussed for
non-resonant pumping in Ref.~\cite{wouters10}). Therefore, taking the
difference between opposite momenta across the vortex core, we have
obtained the average angular velocity to be $\sim 0.2$
$\mu$m$/$ps. This value sets a lower bound for the critical velocity,
which depends on polariton densities and, for small pumping powers,
could be the cause of short lasting times for the polariton vorticity.

The sequence in Fig.~\ref{fig:persi} demonstrates that the vortex
remains steady as a persisting metastable state for times much longer
than the extra population created by the probe pulse and eventually
gets imprinted in the steady state of the OPO signal. This is revealed
by the strong contrast of the fork in the interference images for as
long as the core remains within the condensate area. In the opposite
scenario---dissipation of angular momentum as for classical
fluids---the interference images would show low contrast in the vortex
core region, indicating a mixture of the population which has
undergone dissipation and the one still carrying the quantised angular
momentum.

In our experiments we have measured an average size of the core radius
between $4$ and $5$ $\mu$m. This is in good agreement with a
theoretical estimate of the healing length $\xi \simeq \pi/\sqrt{2
  m_{LP} \delta E}$, where $m_{LP}$ is the lower polariton mass and
$\delta E$ is the lower polariton blue-shift, determined by the
polariton-polariton interaction strength.

%
\section*{Numerical simulations}
%
We have numerically simulated the experiments by making use of
mean-field two-component Gross-Pitaevskii equations for the coupled
cavity and exciton fields with external pump and decay; see the Method
section for details. We have chosen parameters close to the
experimental ones, finding first the conditions for OPO. In the
simulation of Fig.~\ref{fig:thfi1}, we consider a pump with strength
$f_p=1.24 f_p^{\text{(th)}}$ above OPO threshold, and once the steady
state is reached (left panel), we turn on the probe carrying a vortex
$m=1$, resonant with the signal for \unit{2}\pico\second\ only. We
observe a gain of the signal (central panel) for around
\unit{10}\pico\second\ which is followed by a transient time during
which the imprinted vortex drifts around inside the signal. For the
simulation shown in Fig.~\ref{fig:thfi1}, the transient lasts for
about \unit{30}\pico\second, after which the vortex settles into a
metastable solution lasting around \unit{400}\pico\second. In
addition, we also find stable steady-state (i.e., infinitely lived)
vortex solutions, analogous to the ones reported in
Ref.~\cite{whittaker2007}.  Such a solution does not always exist and
it strongly depends on the pumping conditions: in other cases, during
the transient period, the excited vortex either spirals out of the
signal or recombines with an antivortex forming at the edge of the
signal. There are also cases where, instead, the imprinted vortex
settles into a metastable state, which can last several hundred
picoseconds as shown in Fig.~\ref{fig:thfi1}, and then starts drifting
again. We have also analysed the dependence of the vortex solutions on
the probe intensity and found that for steady-state vortex solutions
there is no dependence on the probe intensity, which can only affect
the duration of the transient period. However, for metastable
solutions, we found a threshold in the probe power, below which the
vortex does not get imprinted into the signal anymore. Thus the
results of the theoretical simulations are in excellent qualitative
agreement with the experimental observations.

\begin{figure}[htbp]
\begin{center}
\includegraphics[height=0.83\textheight,width=.9\linewidth,angle=0]{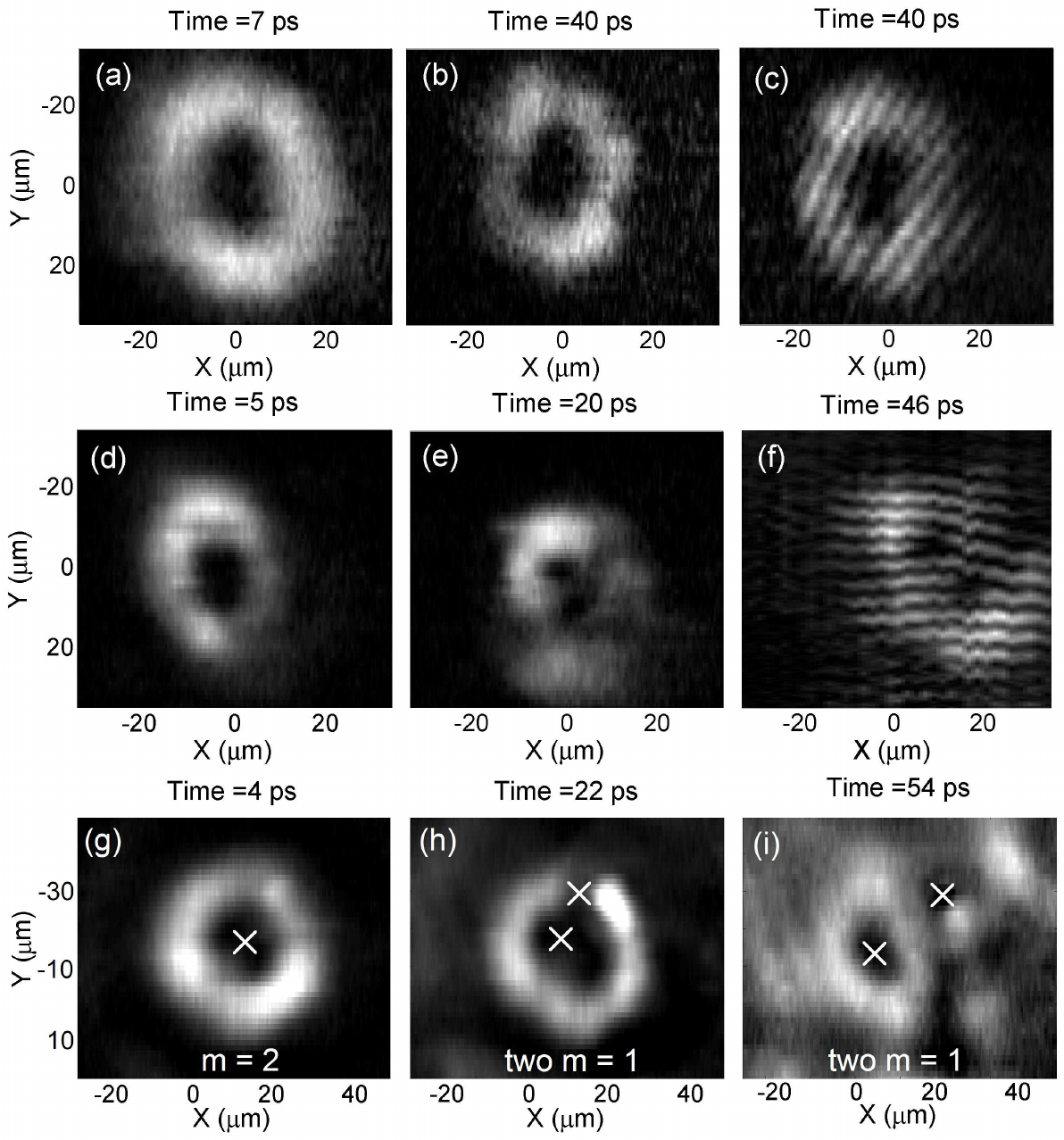}
\end{center}
\vspace{-3cm}
\caption{Time evolution of a $m=2$ vortex TOPO signal excited close to
  zero momentum, $\vect{k}=0$, (a-c) for which neither motion nor
  splitting of the vortex could be detected. However for a signal at
  $\vect{k}_s$ pointing to the right (d-f), we can detect the $m=2$
  vortex both moving and splitting into two single quantised $m=1$
  vortices. In the last panel of the first two rows we plot the
  interference images corresponding to a late time. In both
  experiments the vortex is only present in the TOPO population. On
  the contrary, when the vortex is imprinted into the OPO state, then
  the situation is as shown in the images (g-i). Here the splitting
  appears immediately after the angular momentum is transfered to the
  steady state ($\approx\unit{20}\pico\second$) and only one of the
  two vortices of $m=1$ survives, the other being expelled from the
  condensate in the first \unit{40}\pico\second.}
\label{fig:splitt}
\end{figure}
\begin{figure}[bhpt]
\begin{center}
\includegraphics[width=.65\linewidth]{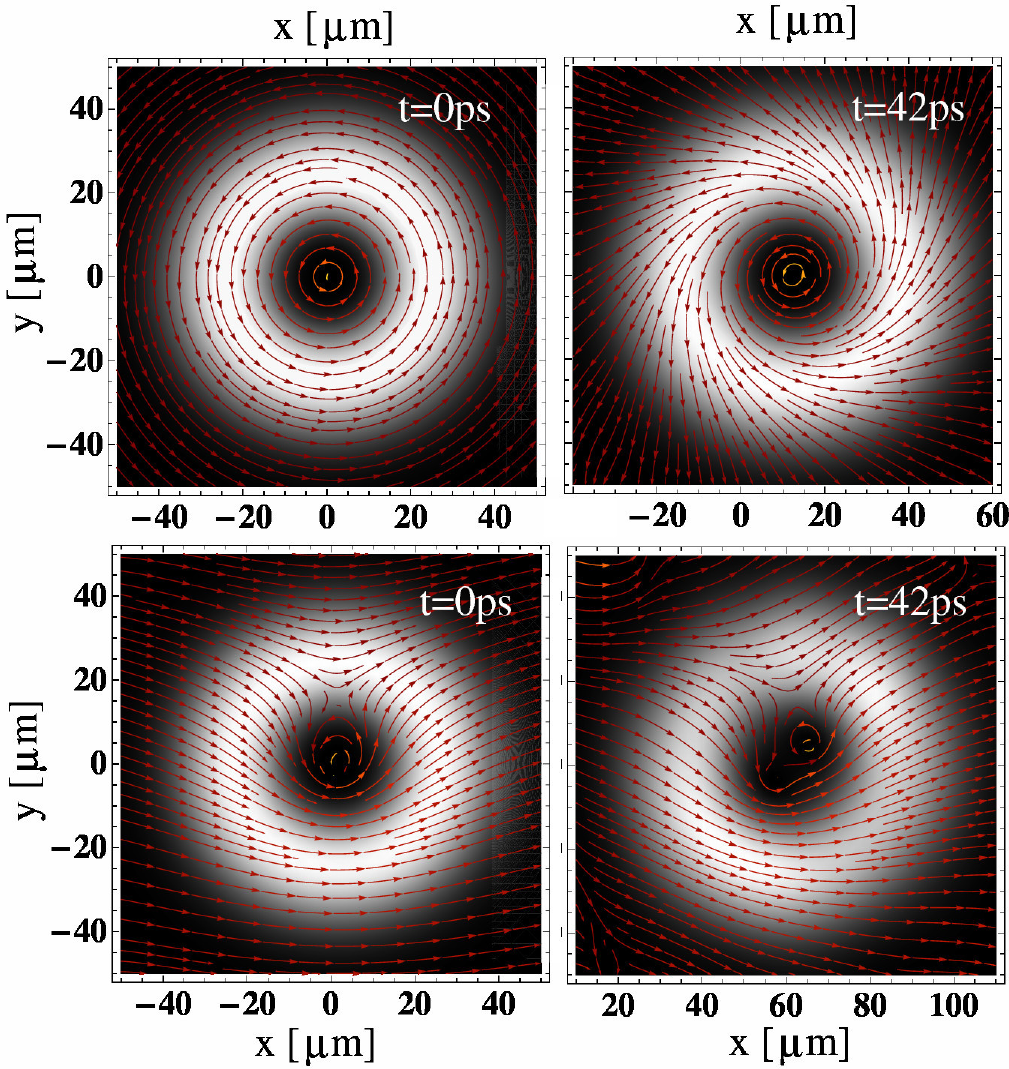}
\end{center}
\caption{Calculated TOPO signal emission for a $m=2$ triggering probe
  at $\vect{k}_{pb} = \unit{0.1}\micro\meter^{-1}< \vect{k}_{pb}^{cr}$
  (first row) and at $\vect{k}_{pb} = \unit{0.7}\micro\meter^{-1} >
  \vect{k}_{pb}^{cr}$ (second row) at the arrival of the probe ($t=0$)
  and \unit{42}\pico\second\ after. The supercurrents are plotted in
  the frame of the group velocity of the moving vortex. In the second
  row, the net current felt by the $m=2$ vortex causes it to split.}
\label{fig:thdou}
\end{figure}
%
\section*{Doubly quantised vortices}
%
A second experiment is aimed at investigating the stability of doubly
quantised vortices in different regimes. We perform both experiments
and numerical simulations in conditions similar to the ones previously
described, but now considering a pulsed probe carrying a doubly
quantised vortex $m=2$. We demonstrate three distinctive behaviours
(see Fig.~\ref{fig:splitt}) by repeating the experiments under
different conditions. For cases in which the vortex lasts as long as
the TOPO population is present, we observe that when the vortex is
excited on a static signal centered at $\vect{k}=0$, it does not split
within its lifetime [Fig.~\ref{fig:splitt} (a-c)]. However, exciting
the signal with a finite momentum, thus making it moving inside the
pump spot, we observe the doubly quantised vortex splitting into two
singly quantised vortices, as shown in Fig.~\ref{fig:splitt} (d-f).

Our theoretical analysis gives the same result: a TOPO vortex (below
the OPO threshold) is stable when it is either at rest or moves below
a critical velocity, and splits otherwise (see Fig.~\ref{fig:thdou}).
In order to explain this result, we have analysed the supercurrents
characterizing the signal. We found that, below threshold, these
supercurrents are a superposition of the vortex currents with a net
current corresponding exactly to the momentum $\vect{k}_{pb}$ at which
the probe has been injected. In addition, we have evaluated the group
velocity $\vect{v}_g$ at which the triggered signal (carrying the
vortex) moves in real space. We found that $\vect{v}_g$ increases
linearly with $\vect{k}_{pb}$ up to a critical value,
$\vect{k}_{pb}^{cr}$, and sublinearly above. Therefore, we expect that
in the frame of the moving vortex, there is no net current in the
linear regime, while, in the sublinear regime, the vortex feels a net
current which causes it to split (see Fig.~\ref{fig:thdou}). Indeed,
we observe that the $m=2$ vortex splits exactly when is injected at or
above $\vect{k}_{pb}^{cr}$. Note that a TOPO signal is a decaying
state, therefore the $m=2$ vortex can be stable only within its
lifetime.

In the case when, instead, the vortex is also imprinted into the OPO
steady-state, we observe that the $m=2$ vortex splits, one $m=1$
vortex is quickly expelled outside the signal, while the other vortex
stabilizes and persists, as is also the case in the experiment, see
Fig.~\ref{fig:splitt} (g-i). Numerical simulations show that above OPO
threshold the structure of the currents in the signal is complex and
thus a stationary vortex will always feel a net current. This is the
reason why above OPO we always observe splitting of the $m=2$ vortex.


\section*{Acknowledgements}
%
We are grateful to D.~Whittaker, J.~J.~Garc\'ia-Ripoll,
P.~B.~Littlewood and J.~Keeling for stimulating discussions. This work
was partially supported by the Spanish MEC (MAT2008-01555 and
QOIT-CSD2006-00019) and the CAM (S2009/ESP-1503).
DS and FMM acknowledge financial support from the
Ram\'on y Cajal program. GT thanks the FPI scholarship from the
Ministry of Education. We would like to thank TCM group (Cavendish
Laboratory, Cambridge, UK) for the use of computer resources.
Requests should be addressed to D.S. for experimental details and
to F.M.M. for theoretical matters.

\section*{Author contributions}
%
D.S., G.T. and M.B. carried out the experiments. F.M.M. and M.H.S.
performed the theoretical simulations. L.M. provided the holograms
for getting vortex excitation and A.L. and J.B. fabricated the samples.
All the authors analysed the results, discussed the underlying physics and
contributed to the manuscript.

\section{Methods}
%
\subsection{Experiments}
%

The sample studied is a $\lambda/2$ AlAs microcavity with a
\unit{20}\nano\meter\ GaAs quantum well placed at the antinode of the
cavity electromagnetic field. The cavity is formed by two high
reflective Bragg mirrors of $25$ pairs at the bottom and $15.5$ on the
top of the structure. The pump is obtained with a cw Ti-Sapphire
laser, resonantly exciting the LPB close to the inflection point at
$9^{\circ}$. The spot size is of $\approx\unit{100}\micro\meter$.

The experiments are performed at a cryogenic temperature of
\unit{10}\kelvin\ and using a very high numerical aperture lens
($0.6$) so that the sample could be accessed by angles as large as
$25^{\circ}$ and the photoluminescence (PL) can be simultaneously
collected from the signal state in the near as well as the far
field. The PL was collected through a \unit{0.5}\meter\ spectrometer
into a streak camera, working in synchroscan mode, with
\unit{4}\pico\second\ time resolution, allowing for energy---as well
as time-resolved---images. In order to reach high time resolution, all
the images where obtained by filtering the signal in the far field
without spectrally resolving the emission of the OPO
states. Differently from our TOPO experiment~\cite{ballarini09}, here
we excite and trigger with the probe at the signal state in a pumping
regime above threshold.

The vortex probe state is prepared by scattering a Ti:Sapphire pulsed
laser Gaussian beam in an hologram with a single or double fork-like
dislocation.  This gives rise to a first order Laguerre-Gauss beam
with a winding number either $m=1$ or $m=2$, respectively. The probe
beam is focused at the center of the cw pump in resonance with the
signal emission energy and at around $\vect{k}=0$ for most of the
cases, except when a finite velocity is given to the vortex state. In
this latter case the probe is arriving on the sample with a finite
angle, between one or two degrees, so that the vortex can have a
finite velocity showing splitting of the $m=2$ state. The probe is
focussed in a region of $\approx\unit{25}\micro\meter$ having a power
set to be below the intensity of the signal. Moreover, considering the
wide energy spread given by its fast duration (\unit{2}\pico\second)
the amount of power getting into the cavity is estimated to be between
1/10 and 1/5 of the signal emission. Every picture is the result of an
average over many shots, as single shot measurements would give a too
low signal to noise ratio.

A realization of optically transferring orbital angular momentum to atomic BECs, using Laguerre-Gauss
beams, as in our experiments, was performed using a two-photon
stimulated Raman process ~\cite{andersen06}. $m=2$ were generated in a
BEC of sodium atoms but its stability was not analysed.

\subsection{Theory}
%
The dynamics of amplitudes and phases of TOPO is analysed using
two-component Gross-Pitaevskii equation with external pumping and decay
for the coupled cavity and exciton fields
$\psi_{C,X} (\vect{r},t)$ ($\hbar=1$):
\begin{equation}
  i\partial_t \begin{pmatrix} \psi_X \\ \psi_C \end{pmatrix} =
  \begin{pmatrix} \omega_X -i \kappa_X + g_X|\psi_X|^2& \Omega_R/2 \\
  \Omega_R/2 & \omega_C -i \kappa_C \end{pmatrix} \begin{pmatrix}
  \psi_X \\ \psi_C
  \end{pmatrix}
  +\begin{pmatrix} 0 \\ F_p + F_{pb} \end{pmatrix} \; .
\label{eq:mode2}
\end{equation}
Since the exciton mass is four orders of magnitude larger than the
photon mass, we neglect the excitonic dispersion and assume a
quadratic dispersion for the cavity photon, $\omega_C=\omega_C(0)
-\nabla^2/2m_C$. The fields decay with rates $\kappa_{C,X}$.
$\Omega_R$ is the photon-exciton coupling. The cavity field is driven
by an external cw pump field,
\begin{equation*}
  F_p(\vect{r},t) = f_p
  e^{-\frac{|\vect{r}-\vect{r}_p|^2}{2\sigma^2_p}} e^{i (\vect{k}_p
    \cdot \vect{r} - \omega_p t)}\; ,
\end{equation*}
while the probe is a Laguerre-Gaussian pulsed beam,
\begin{equation}
  F_{pb}(\vect{r},t) \simeq f_{pb} |\vect{r}-\vect{r}_{pb}|^m
  e^{-\frac{|\vect{r}-\vect{r}_{pb}|^2}{2\sigma^2_{pb}}} e^{i m
    \varphi (\vect{r})} e^{-\frac{(t-t_{pb})^2}{2\sigma^2_{t}}} e^{i
    (\vect{k}_{pb} \cdot \vect{r} - \omega_{pb} t)} \; ,
\label{eq:probe}
\end{equation}
producing a vortex at $\vect{r}_{pb}$ with winding number $m$.  The
exciton repulsive interaction strength $g_X$ can be set to one by
rescaling fields and pump strengths. We solve Eq.~\eqref{eq:mode2}
numerically by using the 5$^{\text{th}}$-order adaptive-step
Runge-Kutta algorithm on a 2D grid.

We take $m_C=2.3 \times 10^{-5} m_0$ and a Rabi splitting $\Omega_R
=\unit{4.4}\milli\electronvolt$ determined experimentally. In the
regime of our experiments, we can neglect the saturation of the dipole
coupling~\cite{ciuti03}. We choose the pumping angle $\vect{k}_p$, the
energy of the pump $\omega_p$ and the pump profile as the experimental
ones.

The calculated images of the signal in Figs.~\ref{fig:thfi1} and
\ref{fig:thdou} are obtained by filtering in momentum space around the
signal momentum, $\vect{k_s}$. We set to zero all the emission in
momentum space aside the one coming from the signal and fast Fourier
transform back to real space. In this way the strong emission coming
at the pump angles is masked out.

\newpage
%
\section{Supplementary material}
%
\begin{figure}[hbpt]
\centering
\includegraphics[width=.8\linewidth,angle=0]{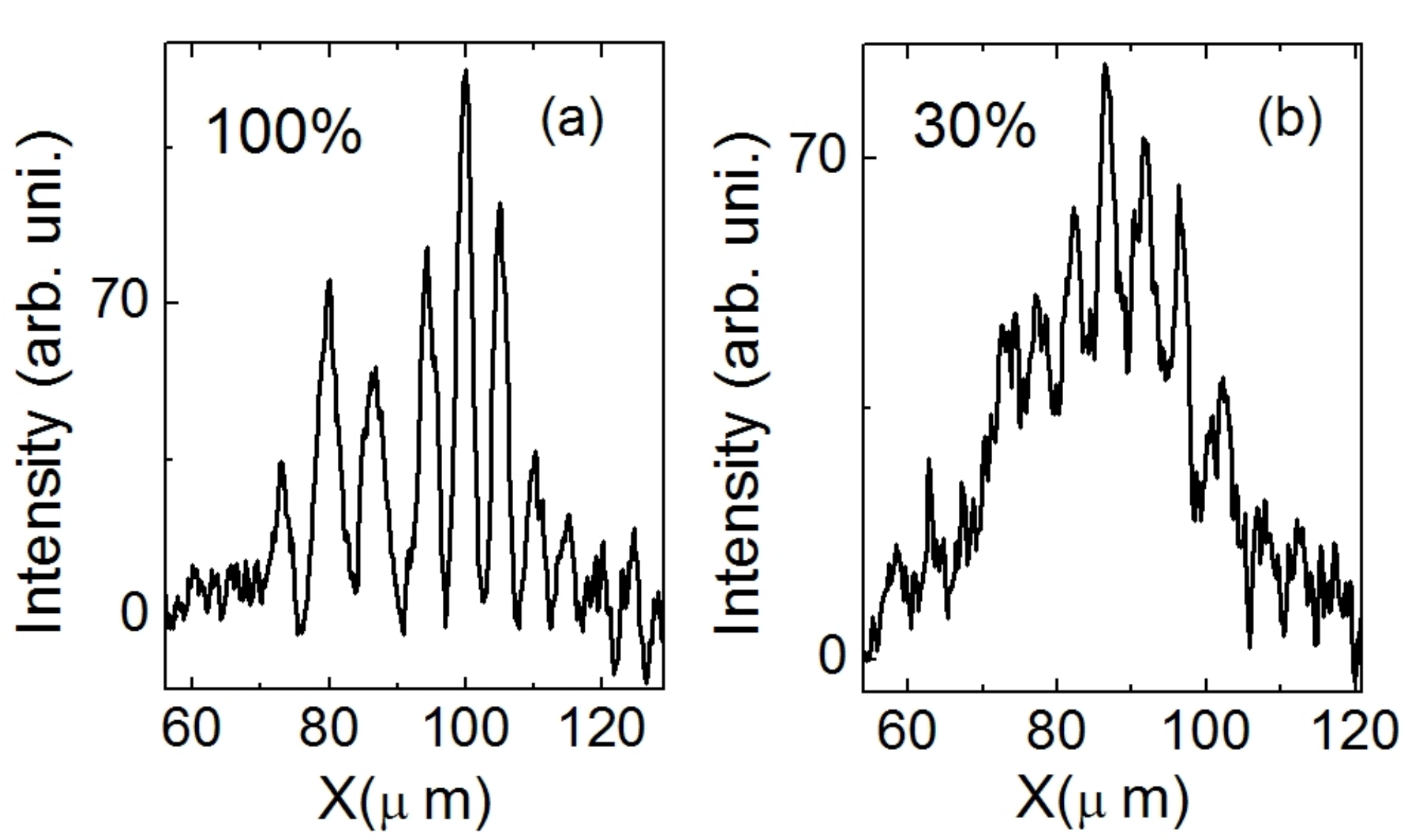}
\caption{Intensity of interference fringes (arbitrary units) of the
  polariton emission after arrival of the probe pulse. The two images
  are taken at different delay times between the polariton signal and
  its mirrored counterpart: (a) \unit{0}\pico\second\ and (b)
  \unit{55}\pico\second\ time delay.}
\label{fig:visib}
\end{figure}

\begin{figure}[hbpt]
\begin{center}
\includegraphics[width=\linewidth,angle=0]{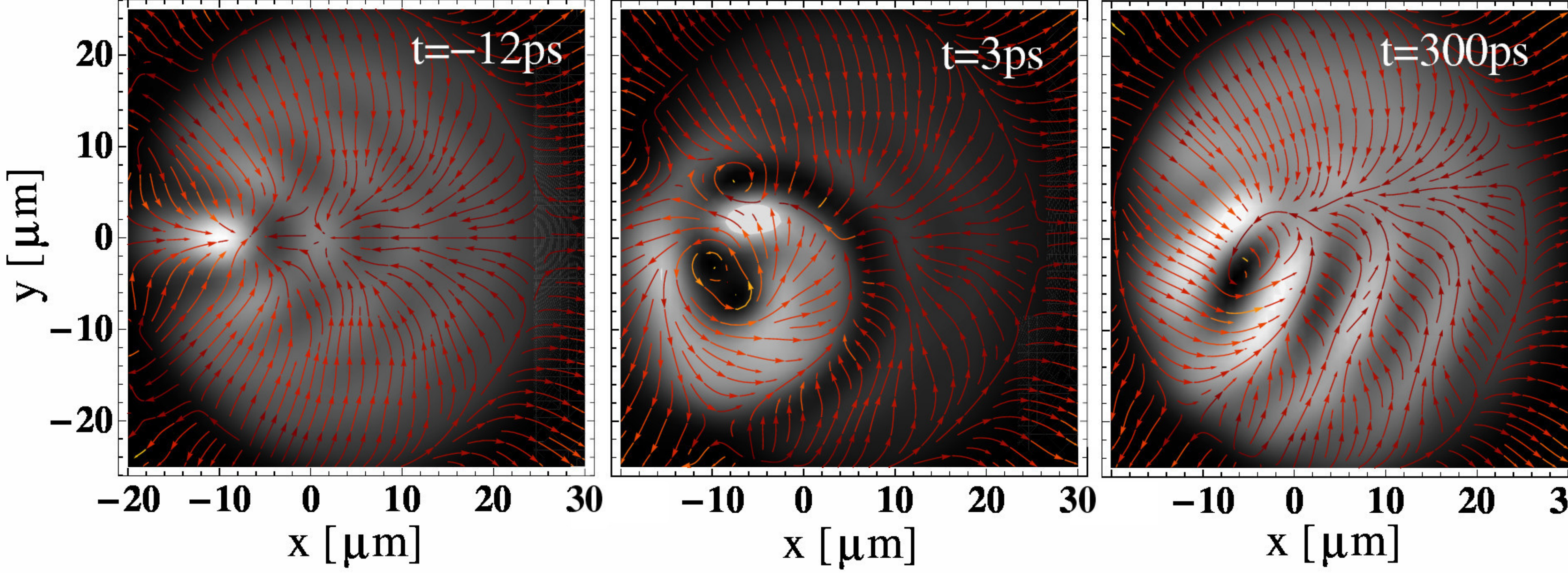}
\end{center}
\caption{Calculated time evolution of the OPO signal emission for a $m=2$
  triggering probe before the arrival of the probe at $t=0$ and after,
  showing the splitting of the imprinted doubly quantised vortex, the
  expulsion from the signal of one $m=1$ vortex and the stabilisation
  of the other $m=1$ vortex into the signal. The supercurrents are
  plotted in the frame of the signal like in the
  Fig.~\ref{fig:thfi1}.}
\label{fig:suth1}
\end{figure}
\begin{figure}[hbpt]
\begin{center}
\includegraphics[width=\linewidth,angle=0]{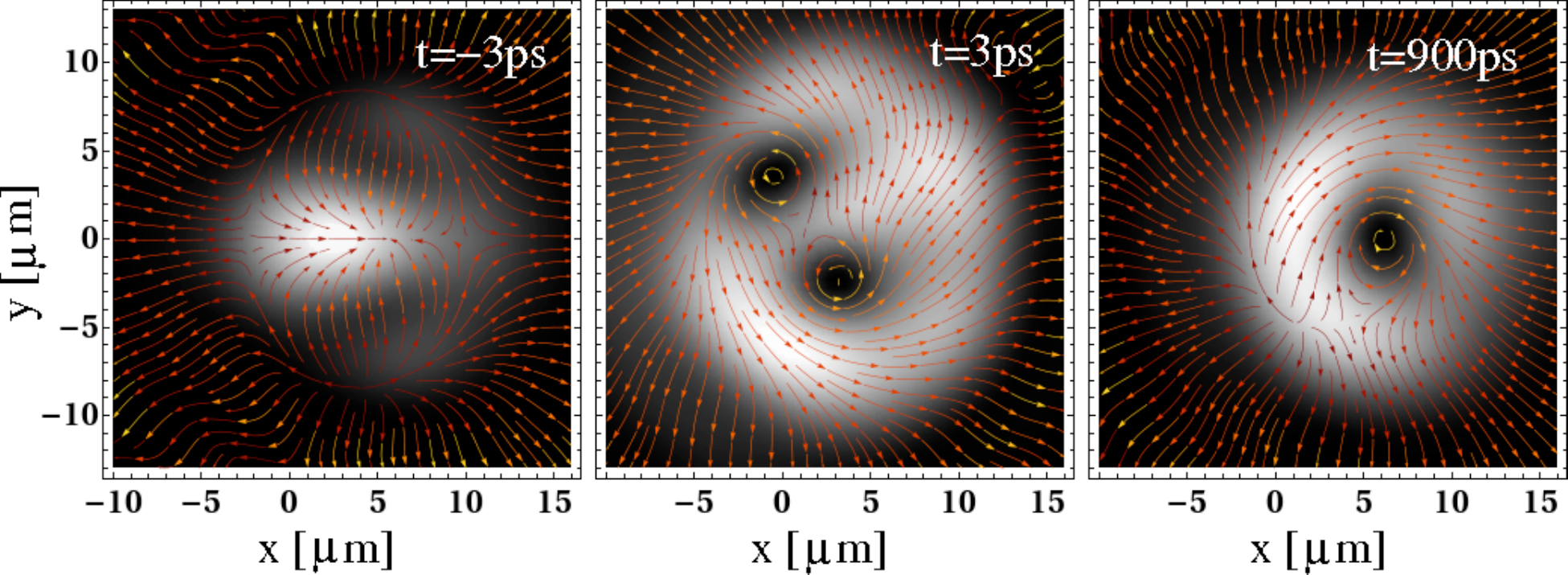}
\end{center}
\caption{Calculated time evolution of the OPO signal emission for a
  $m=2$ triggering probe as in
  Fig.~\ref{fig:suth1} but for smaller spatial size of the signal showing that
  the qualitative behaviour of doubly quantised vortices remains the same for
   smaller signals.}
\label{fig:suth2}
\end{figure}

In Fig.~\ref{fig:visib} we show the intensity of the interference
fringes of the vortex with $m=1$ (Fig.~\ref{fig:persi} of the main
text) obtained with the Michelson interferometer but at different time
delays between the two arms. Immediately after the probe has arrived,
and at zero delay, the fringes exhibits a visibility close to $100\%$,
while it reduces to $30\%$ for a delay of
\unit{55}\pico\second. However the interferences of the vortex core of
Fig.~\ref{fig:persi}---taken always at zero delay time---do not show
any degradation of the visibility even for more than
\unit{100}\pico\second\ after the probe has arrived, demonstrating the
persistence of the polariton vorticity.

In Fig.~\ref{fig:suth1} and \ref{fig:suth2} we show the calculated
time evolution of a doubly quantised vortex in the case when it
imprints into the OPO signal. In this regime, we see the vortex to
split into two singly quantised vortices almost immediately when the
probe arrives, already during the imprinting process. The two vortices
coexist for a short time, after which one is expelled from the
condensate, usually annihilating with an anti-vortex present at the
signal boundary.

\newpage

\newcommand\textdot{\.}

\end{document}